\documentclass[]{aastex63}
\usepackage{amsmath}
\usepackage{bm}

\usepackage{scalerel,stackengine}
\stackMath
\newcommand\reallywidehat[1]{%
\savestack{\tmpbox}{\stretchto{%
  \scaleto{%
    \scalerel*[\widthof{\ensuremath{#1}}]{\kern.1pt\mathchar"0362\kern.1pt}%
    {\rule{0ex}{\textheight}}
  }{\textheight}%
}{2.4ex}}%
\stackon[-6.9pt]{#1}{\tmpbox}%
}

\newcommand{\bk}{{ \bm{k} }}
\newcommand{\bx}{{ \bm{x} }}
\newcommand{\bu}{{ \bm{u} }}
\renewcommand{\bv}{{ \bm{v} }}
\newcommand{\bH}{{ \bm{\Lambda} }}
\newcommand{\sN}{{\mathcal N}}
\newcommand{\sW}{{\mathcal W}}
\newcommand{\del}{{\partial}}
\newcommand{\abs}[1]{{\left\lvert #1 \right\rvert}}

\shortauthors{Lee, Gammie}
\shorttitle{Disks as GRFs}
\begin{document}

\title{Disks as Inhomogeneous, Anisotropic Gaussian Random Fields}
\author{Daeyoung Lee}
\affil{Department of Physics, University of Illinois, 1110 West Green Street, Urbana, IL, 61801; {\tt dl6@illinois.edu}}
\author{Charles F. Gammie}
\affil{Department of Astronomy, University of Illinois, 1002 West Green Street, Urbana, IL, 61801}
\affil{Department of Physics, University of Illinois, 1110 West Green Street, Urbana, IL, 61801; {\tt gammie@illinois.edu}}

\begin{abstract}
	
We model astrophysical disk surface brightness fluctuations as an inhomogeneous, anisotropic, time-dependent Gaussian random field. \replaced{The local covariance is restricted to a particular form, the Mat\'ern covariance, that asymptotes to white noise at large scales and a decaying power-law at small scales.  The field is realized as the solution to a stochastic partial differential equation.}{The field locally obeys the stochastic partial differential equation of a Mat\'ern field, which has a power spectrum that is flat at large scales and falls off as a power law at small scales.} We provide a series of pedagogical examples and along the way provide a convenient parameterization for the local covariance.   We then consider two applications to disks.  In the first we generate a movie of a disk.  In the second, by integrating over a movie of a disk, we generate synthetic light curves and show that the high frequency slope of the resulting power spectrum depends on the local covariance model.  We finish with a summary and a brief discussion of other possible astrophysical applications.

\end{abstract}

\section{Introduction}

Astrophysical disks experience surface brightness fluctuations that can in principle be predicted by numerical solution of well-known governing equations.  Three dimensional physical simulations are computationally expensive, however, and may still be missing important physical processes and be unable to resolve important physical lengthscales.  What if the main features of the fluctuations could be captured in a simpler, easier-to-compute statistical model?  In this paper we consider a model in which the surface brightness fluctuations are treated as a Gaussian random field (GRF).
	
GRFs are widely used in astrophysics to model correlated noise.  In cosmology they are used to model initial conditions \citep[e.g.][]{bar86}.  In studies of quasars the light curve of the unresolved source is commonly modeled as a damped random walk \citep{kel09, mac10}, which is a Gaussian process.  In both cases the field is homogeneous and so a realization can be easily generated by drawing uncorrelated Fourier amplitudes from a Gaussian distribution with variance given by the power spectrum.

In astrophysical disks the correlation length and correlation time likely vary with local radius, possibly by orders of magnitude.  It is not easy to see how Fourier techniques might be generalized to capture this inhomogeneity.  In addition, the surface brightness correlation is likely anisotropic, as for example in the trailing spiral structures that populate flocculent spiral galaxies.  How can one efficiently generate a realization of an inhomogeneous, anisotropic GRF to model fluctuations in astrophysical disks?

In practice we wish to sample the GRF on a finite mesh of points, which is equivalent to sampling a multivariate Gaussian with a prescribed covariance matrix.  For small sets of sample points one might sample directly from the multivariate Gaussian.  A simple but computationally suboptimal way of doing this is to transform to a basis that diagonalizes the covariance matrix (Karhunen-Lo\`eve transformation), draw independent amplitudes, and transform back to the original basis.  A more efficient approach is to use a Cholesky decomposition of the covariance matrix.  This approach is not practical for the large numbers of points considered here. 

One might also generate an inhomogeneous, anisotropic GRF by distorting an initially homogeneous, isotropic GRF through a coordinate transformation \citep{sam92}.  This is, effectively, how weak lensing acts on the microwave background.  The technique is limited to certain geometries and boundary conditions.  Suppose, for example, that one is modeling a flocculent spiral galaxy in which the number of spiral arms changes with radius.  There is no coordinate transformation that can map the associated covariance function onto a rectangular, periodic domain with a homogeneous, isotropic GRF.  

The technique we use here, which is borrowed from geostatistics, generates an inhomogeous, anisotropic field as the solution to a stochastic partial differential equation \citep[SPDE;][]{whi54, whi63, lin11, ful15}.  This method has a well-defined notion of a local covariance function, although the covariance function is limited to a particular - but useful - functional form.  To our knowledge this is the first application of this technique in an astronomical context.  

The plan of this paper is as follows.  In \S \ref{sec:SPDE} we describe the SPDE technique.  In \S \ref{sec:examp} we provide simple examples of anisotropy, inhomogeneity, and time-dependent fields.  In \S \ref{sec:disks} we apply the technique to generation of synthetic movies of differentially rotating disks.  In \S \ref{sec:lightcurve} we explore an application to generating broad band noise for disk light curves. \S \ref{sec:summ} contains a summary and a guide to the main results.

\section{Gaussian Random Fields and SPDEs}
\label{sec:SPDE}

A random field $f$ on a space $X$ is a function such that for every $\bx \in X$, $f(\bx)$ is a random variable. A Gaussian random field (GRF) is a random field such that the joint probability distribution on any set of \replaced{$k$ points is a $k$-dimensional multivariate Gaussian distribution with mean vector $\mu(\bx_i)$ and covariance matrix $C(\bx_i, \bx_j)$. }{$N$ points $\{\bx_1, \ldots, \bx_N\}$ is an $N$-dimensional multivariate Gaussian distribution with mean vector $\mu(\bx_n)$ and covariance matrix $C(\bx_n, \bx_m)$, $n,m \in \{1,\ldots,N\}$. }Thus, a GRF is completely defined by \added{a mean function }$\mu(\bx)$ and \added{a covariance function }$C(\bx, \bx')$\added{, $\bx, \bx' \in X$}. Since $\mu$ can always be subtracted off, we assume without loss of generality that all GRFs have zero mean. 

GRFs have a simple definition and useful analytic properties and have therefore been used as a statistical model for a wide variety of phenomena.  The central limit theorem implies that a superposition of independent, identically distributed random fields is a GRF. Thus, GRFs arise naturally in cosmology, since primordial fluctuations are thought to be the result of an interaction-free (or nearly interaction-free) scalar field during the inflationary period. 

Homogeneous, isotropic GRFs such as those that arise in cosmology are readily generated in a Fourier basis.  Homogeneity implies that the covariance function depends only on $\Delta x$, so
\replaced{\begin{align}
\left\langle \hat{f}(k_i) \hat{f}^*(k_j) \right\rangle &= \int \left\langle f(x) f^*(x') \right\rangle \, e^{i (k_ix - k_jx')} \, dxdx' \nonumber \\ 
&= \int C(s)e^{i (k_i - k_j) x} e^{i k_j s} \,  dxds && \text{$s \equiv  x-x'$} \nonumber \\
&= 2 \pi \hat{C}(k_j) \delta(k_i - k_j)
\end{align}
The Fourier transform of the covariance function $\hat{C}(k)$ is the power spectrum $P(\bk)$.} {\begin{align}
P(k) &= \left\langle \hat{f}(k) \hat{f}^*(k') \right\rangle \nonumber \\
&= \int \left\langle f(x) f^*(x') \right\rangle \, e^{i (kx - k'x')} \, dxdx' \nonumber \\ 
&= \int C(\Delta x)e^{i (k - k') x} e^{i k' \Delta x} \,  dxd\Delta x && \text{$\Delta x \equiv  x-x'$} \nonumber \\
&= 2 \pi \hat{C}(k') \delta(k - k')
\end{align}
where $\left\langle f(x) \right\rangle$ is the expected value of $f(x)$ and $\hat{f}(k)$ is the Fourier transform $\hat{f}(k) = \int f(x) e^{i k x} dx$. Thus, the Fourier transform of the covariance function $C(\Delta x)$ is the power spectrum $P(\bk)$.} Thus one can generate a realization of a homogeneous GRF by drawing independent, normally distributed Fourier amplitudes with variance given by $P(\bk)$.  If the GRF is also isotropic then the power spectrum depends only on $\lvert \bk \rvert$ \citep[e.g.][]{bar86}.  

Inhomogeneity (or non-stationarity)\footnote{The terms homogeneity and stationary are sometimes used interchangeably in literature to describe GRFs. We will use homogeneity as a strictly spatial property and stationarity as a temporal, spatiotemporal, or purely mathematical property. There are two types of stationarity. Strong stationarity requires that the probability distribution be invariant under translations, while weak stationarity requires only that the mean and covariance functions be translation invariant. For GRFs, the two are equivalent.} 
makes the generation of GRFs more difficult because the Fourier modes are no longer delta-correlated. Since the covariance matrix on a rectangular mesh with $N$ points along each of $d$ dimensions has $N^{2d}$ components and Cholesky decomposition of an $M$x$M$ covariance matrix is an $O(M^3)$ operation, generating a GRF by directly sampling the resulting multivariate Gaussian would require $O(N^{3d})$ operations.  This is too costly for even modest grid size.  Various approaches have been developed to make statistical modeling using inhomogeneous GRFs feasible, including low-rank approximations \added{\citep[e.g.][]{cre08} }or covariance tapering\added{ \citep[e.g.][]{fur06}}. 

The method we use here \citep{lin11} takes advantage of a relationship between a particular covariance function known as the Mat\'ern covariance, a stochastic partial differential equation (SPDE), and Gaussian Markov random fields (GMRFs). The Mat\'ern covariance is stationary and isotropic and has the form  
\begin{equation}
C(\bx, \bm{y}) = \frac{\sigma^2}{2^{\nu - 1} \Gamma(\nu)} \left(\frac{\abs{\bx - \bm{y}}}{\lambda}\right)^\nu K_\nu\left(\frac{\abs{\bx - \bm{y}}}{\lambda}\right),
\end{equation}
\added{$\bx, \bm{y} \in \mathbb{R}^d$. } Here $\lambda$ is a scaling parameter, $\nu$ is a differentiability parameter, and $K_\nu$ is the modified Bessel function of the second kind\added{, order $\nu$}.  For small $r/\lambda \equiv \abs{\bx - \bm{y}}/\lambda$, $C_\nu \sim \sigma^2 (1 - A r^{2\nu})$, where $A$ is a constant.  Notice that for $\nu = 1/3$ this matches the covariance of a passive scalar in Kolmogorov turbulence.\footnote{$\nu = 1/2$ matches the covariance for Burgers turbulence.}  For large $r/\lambda$, $C_\nu \rightarrow 0$, that is, the field decorrelates on scales large compared to $\lambda$.  

GMRFs are discrete Gaussian fields (e.g., random fields sampled on a Cartesian lattice) where the probability distribution of the field at a point depends only on its neighbors, that is, the field has the Markov property.  Most numerical representations of Gaussian fields are discrete fields.  \cite{rue02} demonstrated that a Mat\'ern field is well approximated by a GMRF, and \cite{lin11} showed that such a GMRF can be constructed efficiently using SPDEs for certain values of $\nu$.  In what follows we specialize to $\nu = 2 - d/2$, where $d$ is the number of dimensions; a generalization is discussed in the Appendix.

A field $f(\bx)$ with a Mat\'ern covariance for $\nu = 2 - d/2$ can be expressed as a solution to the SPDE 
\begin{equation}\label{eq:spde}
(1 - \lambda^2 \nabla^2) f(\bx) = \sN \sigma \lambda^{d/2} \sW (\bx)
\end{equation}
where $d$ is the number of dimensions, $\sN$ is a normalization constant, $\sigma^2$ is the variance of the field, and $\sW$ is \replaced{Gaussian white noise with unit variance }{a standardized Gaussian white noise process }\citep{whi54, whi63}. The power spectrum for solutions to (\ref{eq:spde}) is 
\begin{equation}
    P_k = \frac{\sN^2 \sigma^2 \lambda^d }{(1 + (k\lambda)^2)^2}.
\end{equation}
Since Gaussian white noise is easy to generate on a lattice, this changes the computational task from manipulating a large covariance matrix to solving a finite difference approximation to an elliptic  partial differential equation.  The numerically obtained solution of the SPDE on a finite grid or irregular lattice is a GMRF that represents the underlying GRF. 

The key advantage of the SPDE method is that it can be readily generalized to inhomogeneous, anisotropic fields at little extra cost. By taking the SPDE as the {\em definition} of an inhomogeneous Mat\'ern field, we can construct anisotropic and non-stationary variants by introducing position-dependence in the parameters, e.g.
\replaced{\begin{equation}\label{spde}
(1 - \nabla \cdot\bH  (\bx) \cdot \nabla) f(\bx) = \sN \sigma(\bx) \left( \mathrm{det}(\bH) \right)^{1/4} \, \sW (\bx)
\end{equation}
where the matrix $\bH (\bx)$ introduces position-dependent anisotropy and correlation lengths, $\sigma^2$ is the local variance of the field, and $\sN$ is a normalization constant. }
{\begin{equation}\label{spde}
(1 - \nabla \cdot\bH  (\bx) \nabla) f(\bx) = \sN \sigma(\bx) \left[ \mathrm{det}(\bH(\bx)) \right]^{1/4} \, \sW (\bx)
\end{equation}
where the matrix $\bH (\bx)$ introduces position-dependent anisotropy and correlation lengths, $\sigma^2$ is the local variance of the field, $\sN$ is a normalization constant, and $\bx \in \mathbb{R}^2$ or $\mathbb{R}^3$ (since $\nu > 0$). Because varying the correlation lengths also changes the variance of the field, the variance at each point is approximately normalized by the factor of $\left( \mathrm{det}(\bH) \right)^{1/4}$. This normalization holds when the correlation lengths varies on scales large compared to the correlation lengths. }

\section{Examples}
\label{sec:examp}

Before applying this model to astrophysical problems we consider a sequence of models demonstrating the ability to model anisotropy, inhomogeneity \citep[see also][]{ful15}, and time dependence in two spatial dimensions. 

The local covariance is controlled by $\bH$, which we parameterize as
\begin{equation}
    \bH  = \lambda_1^2\bu_1\bu_1^\mathrm{T} + \lambda_2^2\bu_2\bu_2^\mathrm{T}
\end{equation}
where $\lambda_1$ and $\lambda_2$ are the correlation lengths along the axes specified by the 2-d spatial unit vectors $\bu_1$ and $\bu_2$ and $\bu_1 \cdot \bu_2 = 0$. Notice that $\left( \mathrm{det}\bH \right)^{1/2} = \lambda_1 \lambda_2$. If the field is homogeneous, the resulting covariance function is  
\replaced{\begin{equation}\label{2dcov}
C(\Delta \bx) = \frac{\sigma^2}{2^{\nu - 1} \Gamma(\nu)} u^\nu K_\nu(u)
\end{equation}
where
\begin{align}\label{2dmet}
u^2 &= \Delta \bx \cdot \bH^{-1} \Delta \bx \nonumber \\
&= \left(\frac{\Delta \bx \cdot \bu_1}{\lambda_1}\right)^2 + \left(\frac{\Delta \bx \cdot \bu_2}{\lambda_2}\right)^2.
\end{align}}
{\begin{equation}\label{2dcov}
C(\Delta \bx) = \frac{\sigma^2}{2^{\nu - 1} \Gamma(\nu)} s(\Delta \bx)^\nu K_\nu \left( s(\Delta \bx) \right)
\end{equation}
where
\begin{align}\label{2dmet}
s(\Delta \bx)^2 &= \Delta \bx \cdot \bH^{-1} \Delta \bx \nonumber \\
&= \left(\frac{\Delta \bx \cdot \bu_1}{\lambda_1}\right)^2 + \left(\frac{\Delta \bx \cdot \bu_2}{\lambda_2}\right)^2.
\end{align}
}
Evidently $\bH^{-1}$ acts as a metric on the space.  

If $\lambda_1 = \lambda_2 = \lambda$ (i.e. $\bH =\lambda^2 \bm{I}$) the SPDE solution is a homogeneous, isotropic GRF.  Anisotropy is introduced by choosing an anisotropy direction $\bu_1$ and setting $\lambda_1 \ne \lambda_2$. Inhomogeneity is introduced by allowing $\bH$ to vary across the domain. 

\begin{figure}[ht]
	\centering
	\includegraphics[width=0.32\linewidth]{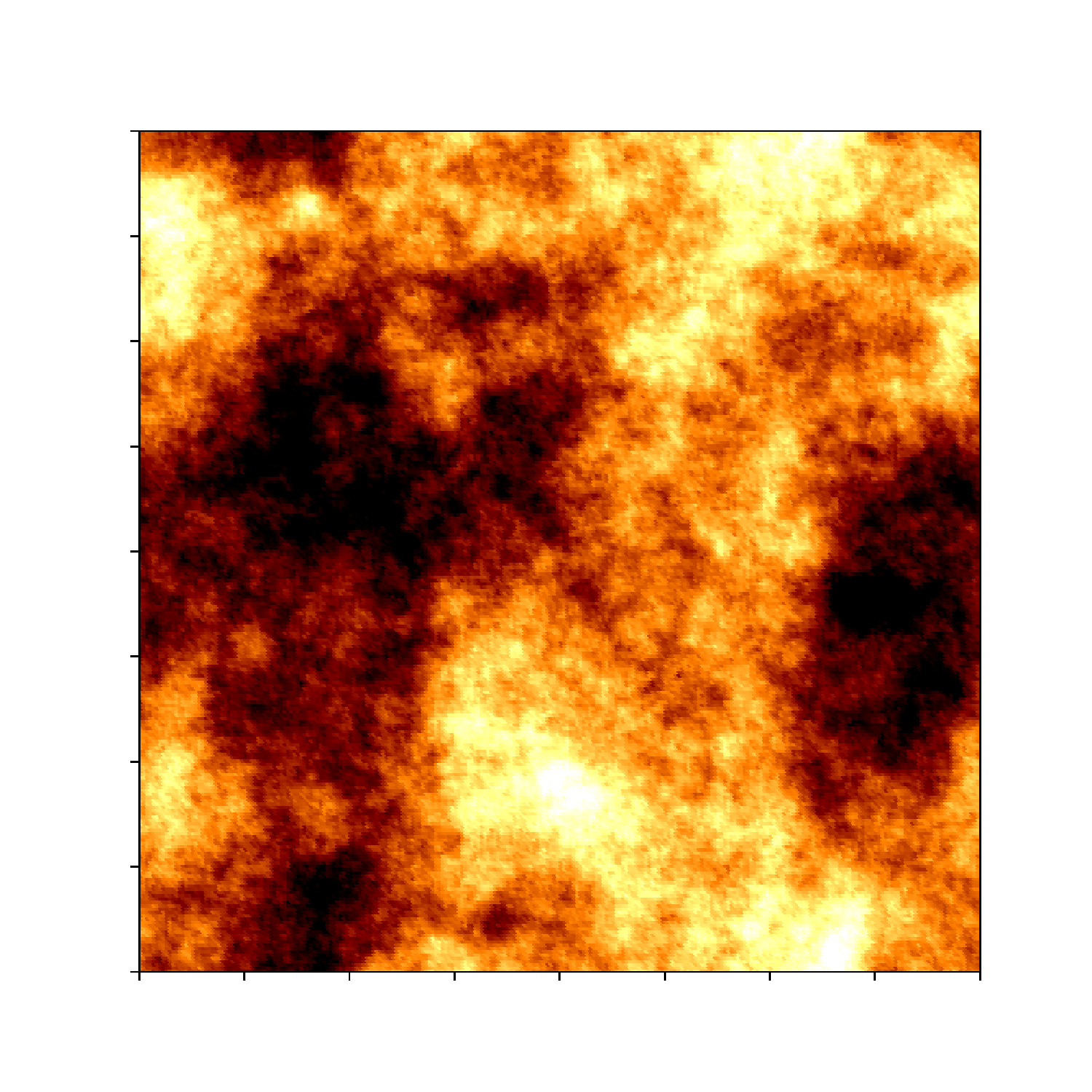}
	\includegraphics[width=0.32\linewidth]{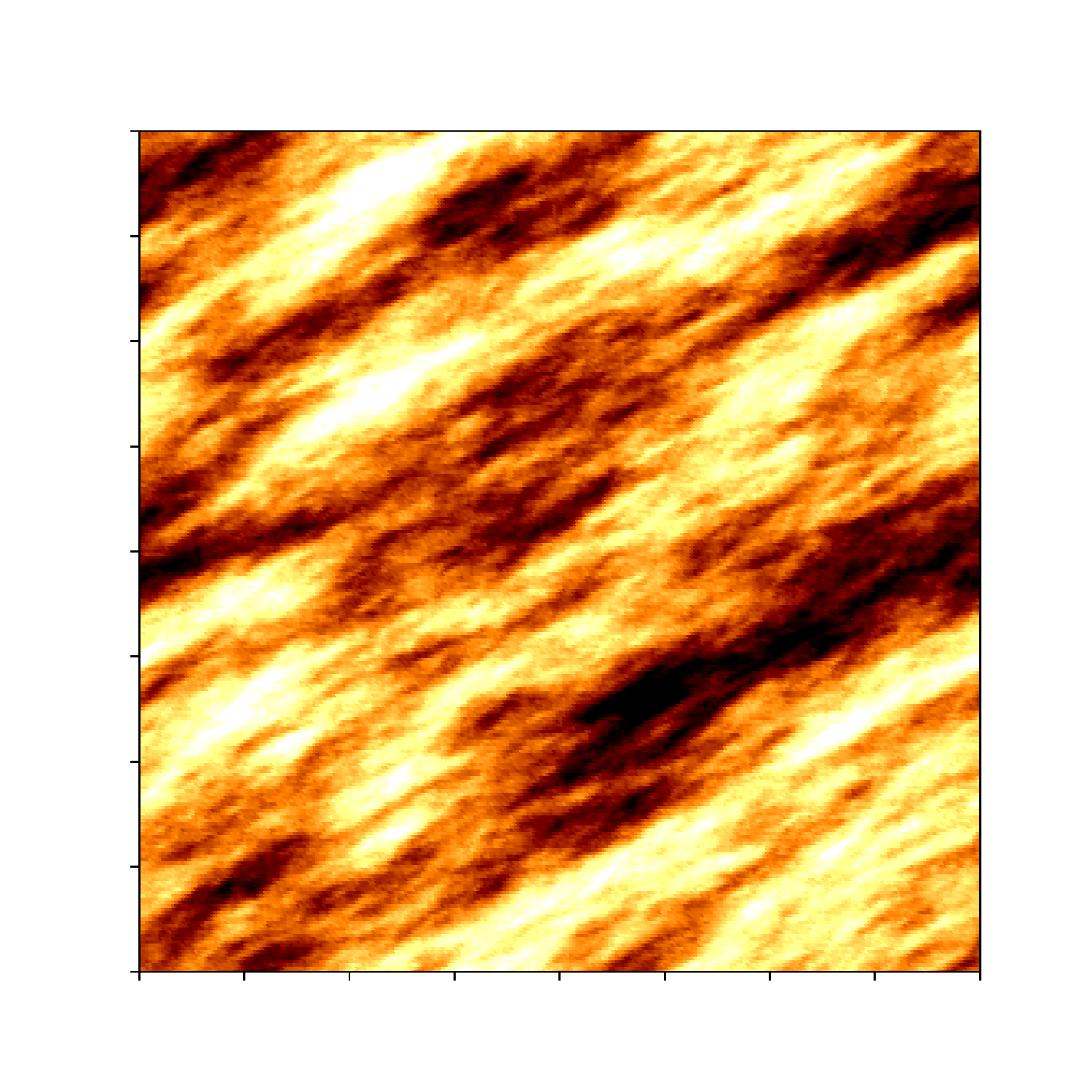}
	\includegraphics[width=0.32\linewidth]{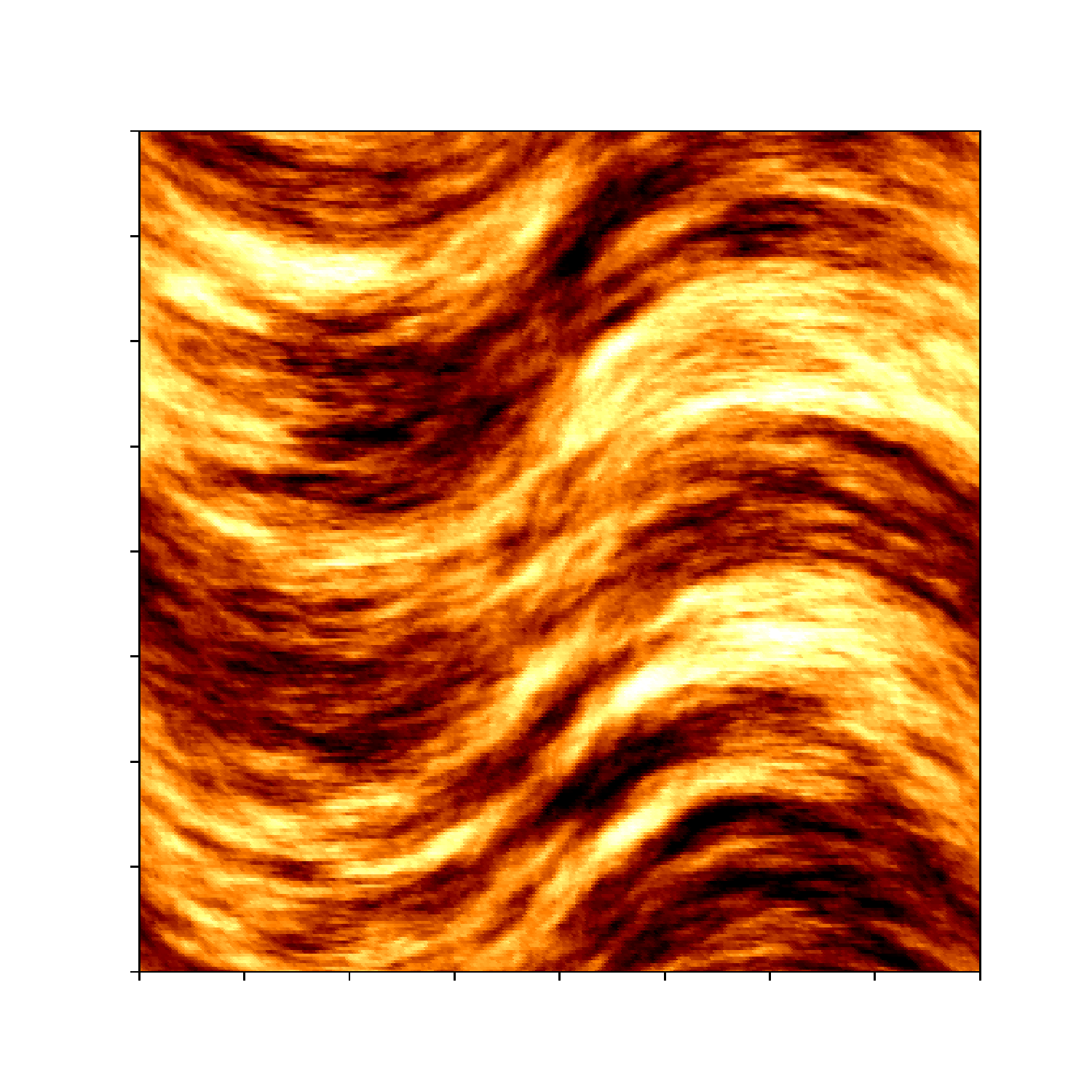}
	\caption{Realizations of a homogeneous, isotropic field (left panel) a homogeneous, anisotropic field (center panel), an inhomogeneous, anisotropic field (right panel). \label{fig:models1}}
\end{figure}

Realizations of isotropic/homogeneous, anisotropic/homogeneous, and anisotropic/inhomogeneous fields are shown in Figure \ref{fig:models1}. Each model was generated on a 256x256 Cartesian grid \added{with periodic boundary conditions}, with $\lambda_1 = 64$ grid spaces. The left model is homogeneous ($\lambda_2 = \lambda_1 = const.$). The center panel is anisotropic with $\bu_1 = (\mathrm{cos}\,\pi/6,\mathrm{sin}\,\pi/6)$ and $\lambda_2 = 0.2\lambda_1$.  The right panel is inhomogeneous with $\bu_1 = (\cos \, \theta(x),\sin \,\theta(x))$ and $\theta(x) = \pi/4 - \pi \abs{x}/L$, where $L$ is the length of the grid and $x \in [-L/2,L/2]$. 

The model can be extended to include time variation by building a \deleted{3D }GRF with two space and one time dimension, and a suitable specification for $\bH$ (the Appendix describes an alternative approach to time dependence).  \replaced{Introducing the velocity field $\bv = (0, v_x, v_y)$ and correlation time $\lambda_0$, we write $u^2 = \Delta \bx \cdot \bH^{-1} \cdot \Delta \bx$, where now
\begin{equation}\label{3dgrf}
    \bH  = \lambda_0^2\bu_0\bu_0^\mathrm{T} + \lambda_1^2\bu_1\bu_1^\mathrm{T} + \lambda_2^2\bu_2\bu_2^\mathrm{T},
\end{equation}
so that 
\begin{equation}\label{3dmet}
    u^2 = \left( \frac{\Delta t}{\lambda_0} \right)^2 + \left( \frac{(\Delta \bx - \bv \Delta t) \cdot \bu_1}{\lambda_1} \right)^2 + \left( \frac{(\Delta \bx - \bv \Delta t) \cdot \bu_2}{\lambda_2}\right )^2.
\end{equation}
where $\bu_0 = (1, v_x, v_y)$, $\bu_1 = (0, \mathrm{cos}\,\theta, \mathrm{sin}\,\theta)$, and $\bu_2 = (0, -\mathrm{sin}\,\theta, \mathrm{cos}\,\theta)$.  Notice that while $\bu_1$ and $\bu_2$ are still unit vectors, $\bu_0$ is not. Nevertheless, $\left( \mathrm{det}\bH \right)^{1/2} = \lambda_0 \lambda_1 \lambda_2$, as in the 2D case. With this parametrization, $\lambda_1$ and $\lambda_2$ are the correlation lengths along the major and minor spatial axes in a single timeslice, and $\lambda_0$ is the correlation time along the shear flow. }
{Here, the vector $\bx = (t,x,y)$. Introducing the velocity field $\bv = (0, v_x, v_y)$ and correlation time $\lambda_0$, we write $s^2 = \Delta \bx \cdot \bH^{-1} \cdot \Delta \bx$, where now
\begin{equation}\label{3dgrf}
    \bH  = \lambda_0^2\bu_0\bu_0^\mathrm{T} + \lambda_1^2\bu_1\bu_1^\mathrm{T} + \lambda_2^2\bu_2\bu_2^\mathrm{T},
\end{equation}
so that 
\begin{equation}\label{3dmet}
    s^2 = \left( \frac{\Delta t}{\lambda_0} \right)^2 + \left( \frac{(\Delta \bx - \bv \Delta t) \cdot \bu_1}{\lambda_1} \right)^2 + \left( \frac{(\Delta \bx - \bv \Delta t) \cdot \bu_2}{\lambda_2}\right )^2.
\end{equation}
where $\Delta \bx = (\Delta t, \Delta x, \Delta y)$, $\bu_0 = (1, v_x, v_y)$, $\bu_1 = (0, \mathrm{cos}\,\theta, \mathrm{sin}\,\theta)$, and $\bu_2 = (0, -\mathrm{sin}\,\theta, \mathrm{cos}\,\theta)$.  Notice that while $\bu_1$ and $\bu_2$ are still unit vectors, $\bu_0$ is not. Furthermore, $\bu_0$ is not orthogonal to $\bu_1$ or $\bu_2$. Nevertheless, $\left( \mathrm{det}\bH \right)^{1/2} = \lambda_0 \lambda_1 \lambda_2$. As in the 2D case, $\bH$ acts as a metric. In particular, along a spatial slice $\Delta \bx = (0, \Delta x, \Delta y)$, the form of $r$ is identical to the 2D case, and thus $\lambda_1$ and $\lambda_2$ are the correlation lengths along the major and minor spatial axes at any point in time. On the other hand, setting $\Delta \bx = (\Delta t, v_x \Delta t, v_y \Delta t) = \Delta t \bu_0$, we have $s^2 = \left( \Delta t / \lambda_0 \right)^2$. Thus $\lambda_0$ corresponds to the correlation time following the flow. }

Figure 2 shows an example that takes the homogeneous, anisotropic model from above and generates a time dependent model with a velocity field in the $-x$ direction, advecting it to the left. 

\begin{figure}[hb]
	\begin{interactive}{animation}{advection.mp4}
		\centering
		\includegraphics[width=0.19\linewidth]{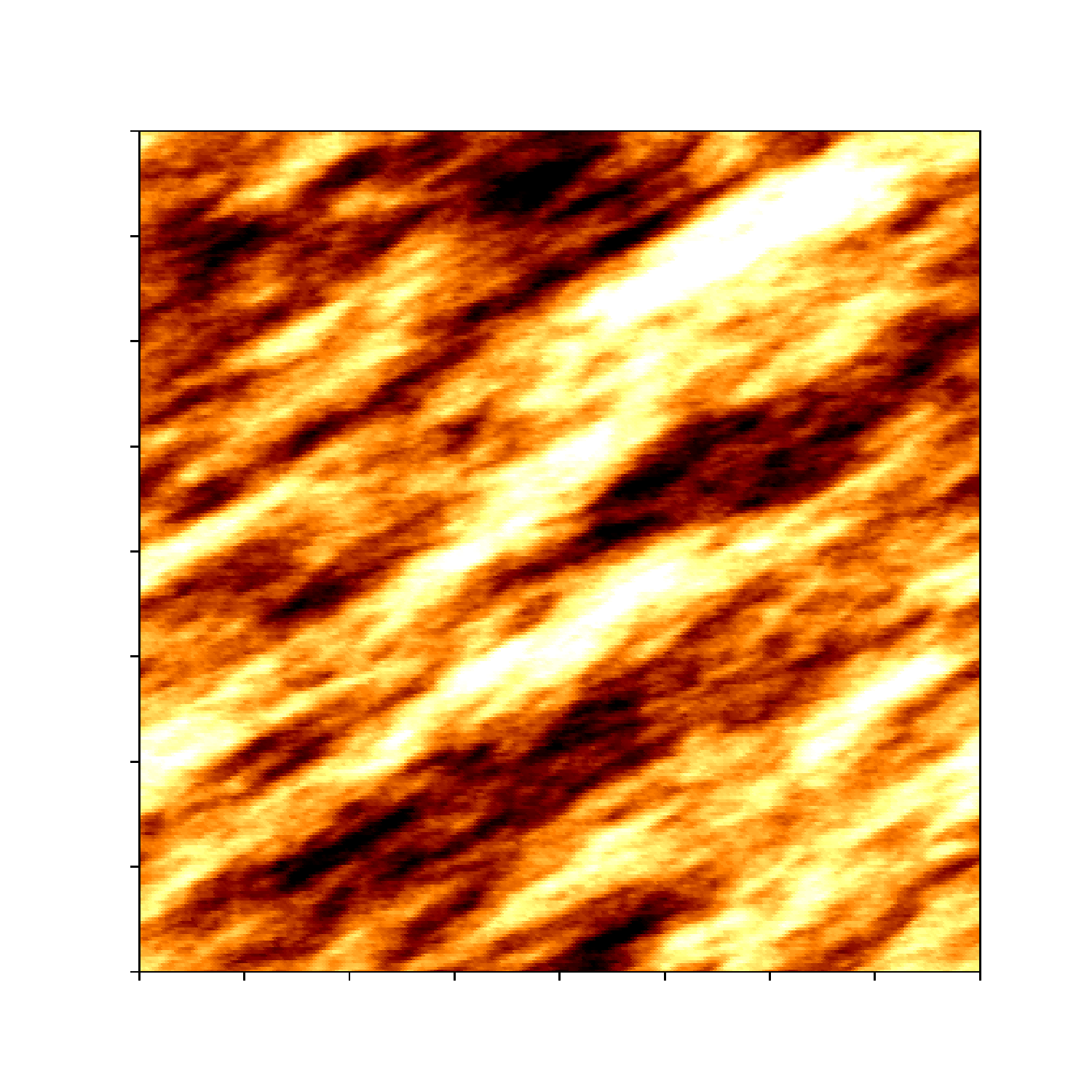}
		\includegraphics[width=0.19\linewidth]{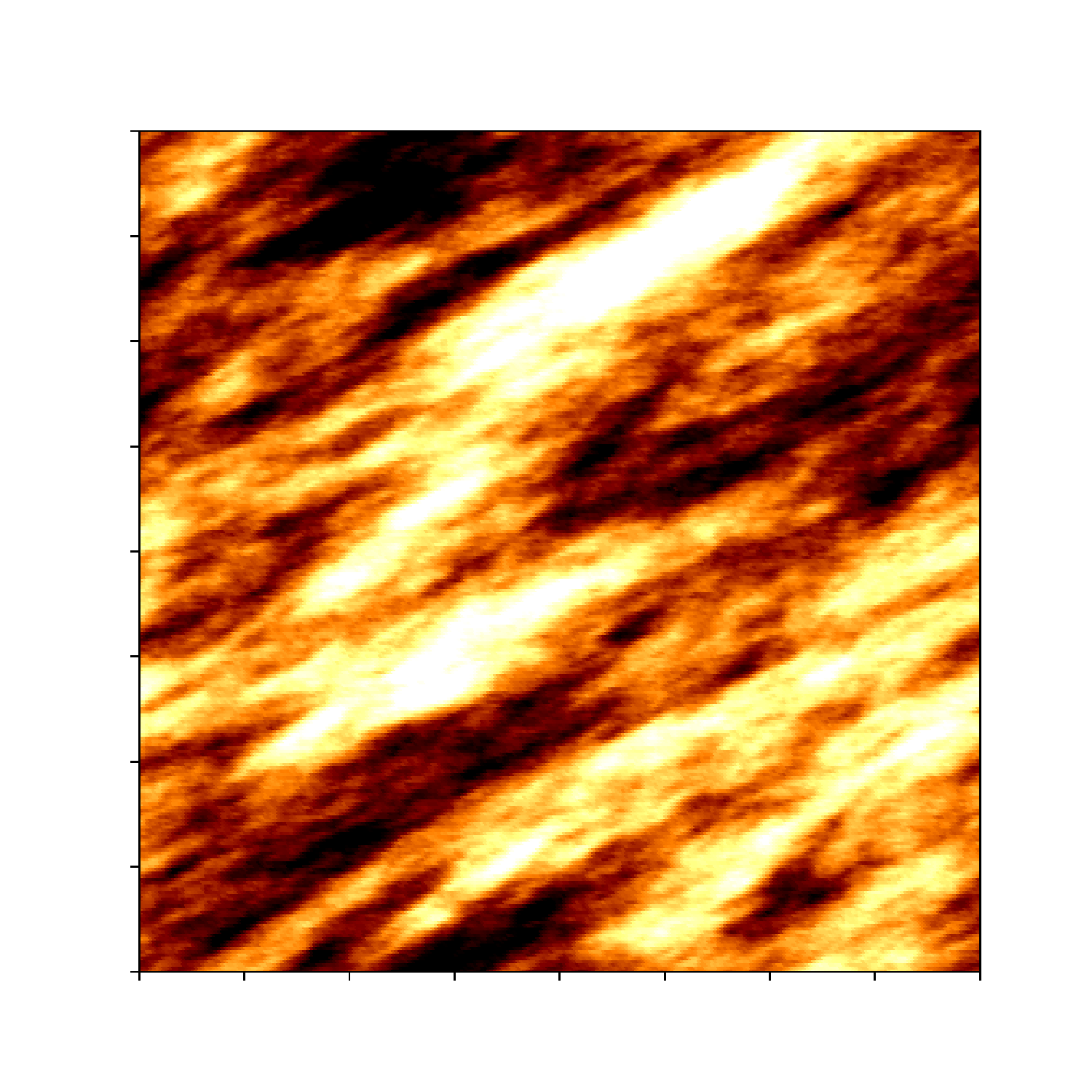}
		\includegraphics[width=0.19\linewidth]{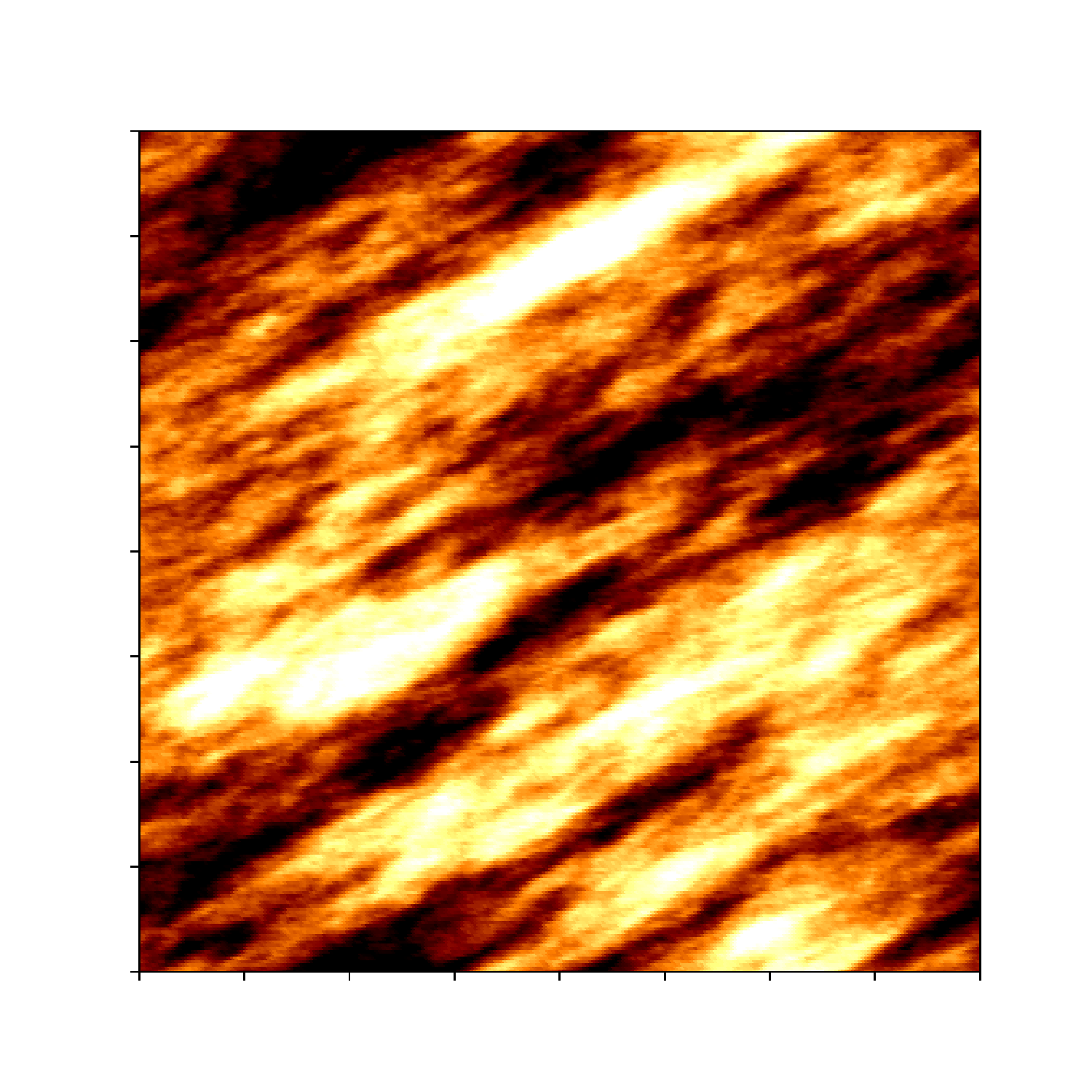}
		\includegraphics[width=0.19\linewidth]{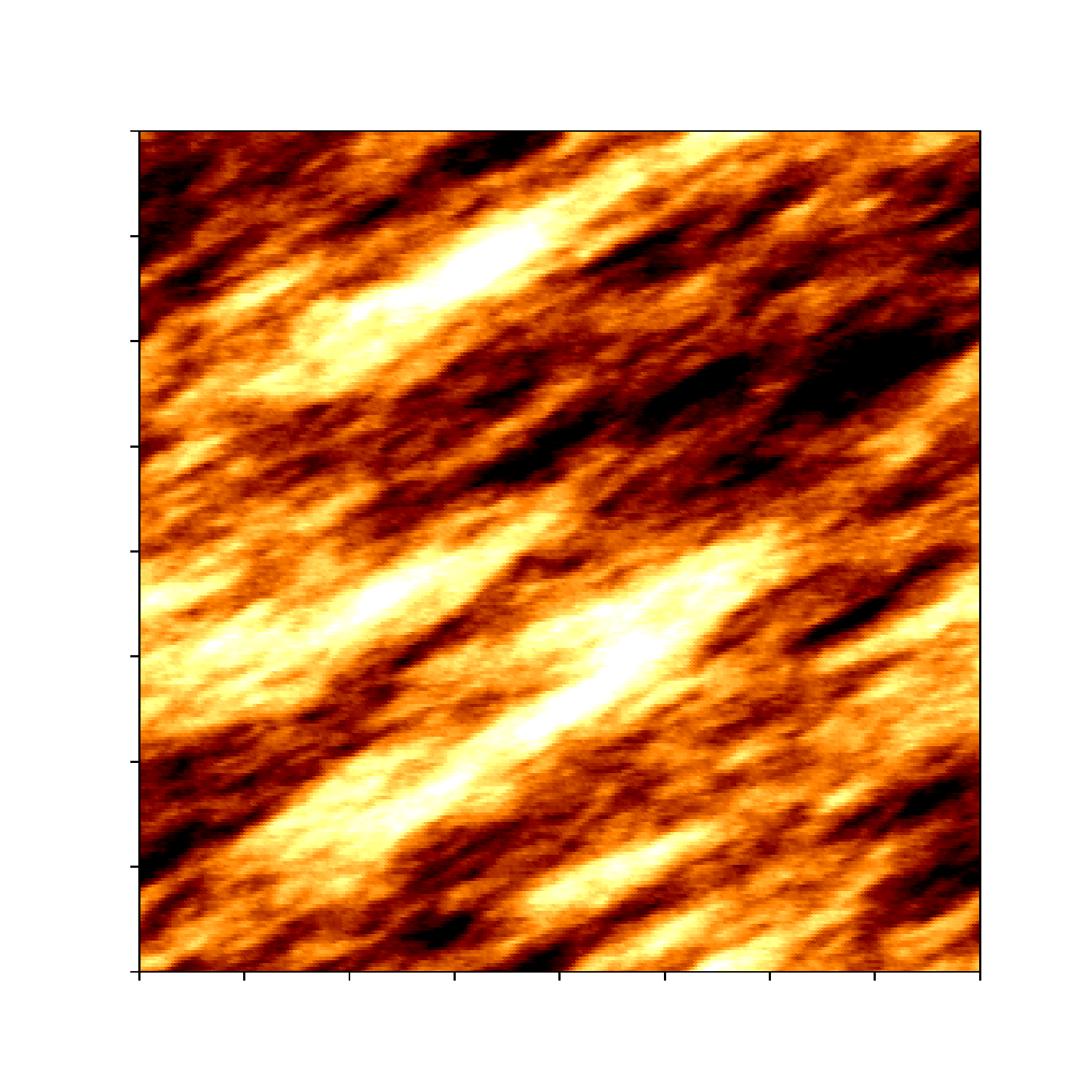}
		\includegraphics[width=0.19\linewidth]{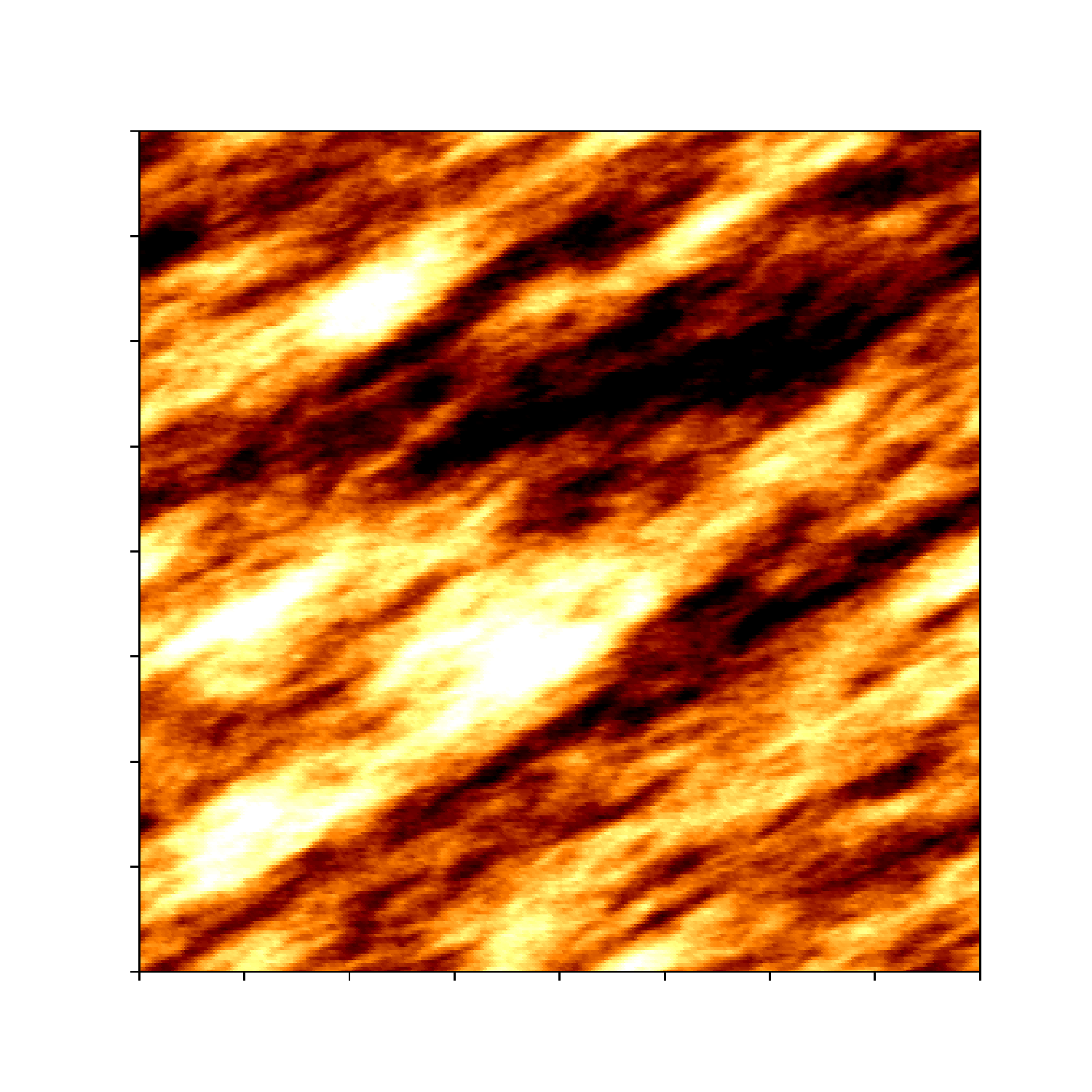} 
	\end{interactive}
	\caption{A realization of an anisotropic, homogeneous field advected to the left by a velocity field. While the field is generated probabilistically, the model encodes a spatiotemporal correlation that is aligned with the velocity field. Thus, across a single correlation time, features appear to be moving across the domain before decorrelating.  \label{fig:animation}}
\end{figure}

Notice that the model (\ref{3dgrf}) is quite flexible and can be used in conjunction with the SPDE technique to introduce velocity fluctuations around any velocity field.  The Appendix describes a generalization to three and more spatial dimensions.

\section{Application to Resolved Disks}
\label{sec:disks}

Next we consider a model for a resolved disk with surface brightness fluctuations.  Possible applications including disk galaxies, disks around young stars (which appear to be notably lacking in turbulence), and planetary rings.  Here we are motivated by Event Horizon Telescope observations of the disk around the black hole in M87 \citep[(EHT)][]{ehta, ehtb, ehtc, ehtd, ehte, ehtf}.  We will consider a simple image-plane model that assigns mean velocities to points on the image.  The model does not include radiative transport, lensing, or other key physical processes.  The local covariances are drawn from numerical simulations of disks, and we adopt the drastic simplification that the mean velocities follow a Keplerian profile.

\subsection{Local Correlations in Disks}

The local covariance of velocity fluctuations in disks can be derived from local model simulations of disks.  \citet{gua09}, for example, give an analytic fit for the covariance function on a plane lying at fixed altitude in a local numerical model of a disk.  Their power spectrum is constant (white noise) at small $k$ and forms a tilted ellipse at large $k$, asymptotically decaying as $k^{-11/3}$. Thus, the local power spectrum can be modeled by 
\replaced{\begin{equation}
P(k) \propto \frac{1}{(1 + \bk \cdot \bH \cdot \bk)^{11/6}}
\end{equation}}
{\begin{equation}
P(\bk) \propto \frac{1}{(1 + \bk \cdot \bH \bk)^{11/6}}
\end{equation}}
which corresponds to a covariance function
\replaced{\begin{equation}
C \propto u^{1/3} K_{1/3}(u)
\end{equation}
where $K_{1/3}$ is a modified Bessel function of the second kind and $u^2 = \Delta \bx \cdot \bH^{-1} \cdot \Delta \bx$, with $\bH$ parameterized as in \S 3. 

We approximate the \citet{gua09} covariance by a $\nu = 1/2$ Mat\'ern field as described in \S 2 and \S 3. This preserves key features of the field: an anisotropic covariance function with steep decay at small scales and decorrelation at large scales. 
}
{\begin{equation}
C \propto s^{1/3} K_{1/3}(s)
\end{equation}
where $K_{1/3}$ is a modified Bessel function of the second kind and $s^2 = \Delta \bx \cdot \bH^{-1} \Delta \bx$. Here, $\bx$ is a 3D spatial vector, and $\bH$ parameterized as a spatially 3D extension of \S 3. 

For simplicity we work in 2D, in the midplane of the disk, and approximate the \citet{gua09} covariance with a $\nu = 1/2$ Mat\'ern covariance. We then extend the spatial covariance to 2+1D as described in \S 2 and \S 3. This preserves key features of the field: an anisotropic covariance function with steep decay at small scales, decorrelation at large scales, and a power spectrum of the form
\begin{equation}
P(\bk) \propto \frac{1}{\left( 1 + \bk \cdot \bH \bk \right)^2};
\end{equation}
here $\bk = (\omega, k_x, k_y)$. }

\subsection{Global Model}
\label{sec:global}

We can now construct a time-dependent global disk model.  First, we produce a realization of the fluctuation field $f$, which we interpret as fractional variation in disk surface brightness, using \eqref{3dgrf}.  

The velocity field is $\bv = \Omega_K \hat{\bm{z}} \times {\bx}$, with $\Omega_K \propto |{\bx}|^{-3/2}$, corresponding to Keplerian rotation.  The major axis of the correlation tensor $\bu_1$ is chosen to lie at a contant $20^\circ$ angle to a circle of constant radius.  This is the opening angle of spiral features in the GRF, and is consistent with  local model results \citep{gua09}.  We let $\lambda_0$, $\lambda_1$, and $\lambda_2$ vary with radius but not azimuth, since the model is on average axisymmetric.  It is natural to take $\lambda_0 \propto 1/\Omega_K$.  The correlation lengths are expected to be proportional to the disk scale height, which is in turn assumed proportional to the local radius (this assumption can be relaxed), so $\lambda_1 \propto r$ and $\frac{\lambda_1}{\lambda_2}=\mathrm{const}$.  Finally, we set $\sigma = 1$.   This completes the specification of $f$.  

We produced a realization by generating Gaussian white noise on a grid and solving the elliptic equation \ref{eq:spde}.  We used the preconditioned conjugate gradient method with a semicoarsening multigrid preconditioner provided by \texttt{hypre}, a library of parallel solvers for linear systems.

To generate a movie of the disk we need to relate $f$ to the surface brightness $\mu$.  The mean surface brightness of the disk is given by an ``envelope function'' $g(r)$.  In this example $x \equiv r_0/r$ and
\begin{equation}
g(r) = x^4 \exp(-x^2).
\end{equation}
This has a ``shadow'' in the middle, like the EHT image of M87, and a surface brightness that drops off as $r^{-4}$ at large radius. The surface brightness is then
\begin{equation}
\mu = g(r)\,\exp\left(\frac{f}{n}\right).
\end{equation}
The control parameter (or function) $n$ controls the fluctuation amplitude (this could also be done when the random field is generated). The envelope function is independent of the GMRF and can be chosen for convenience. 

\begin{figure}[ht]
\begin{interactive}{animation}{realization.mp4}
		\centering
		\includegraphics[width=0.32\linewidth]{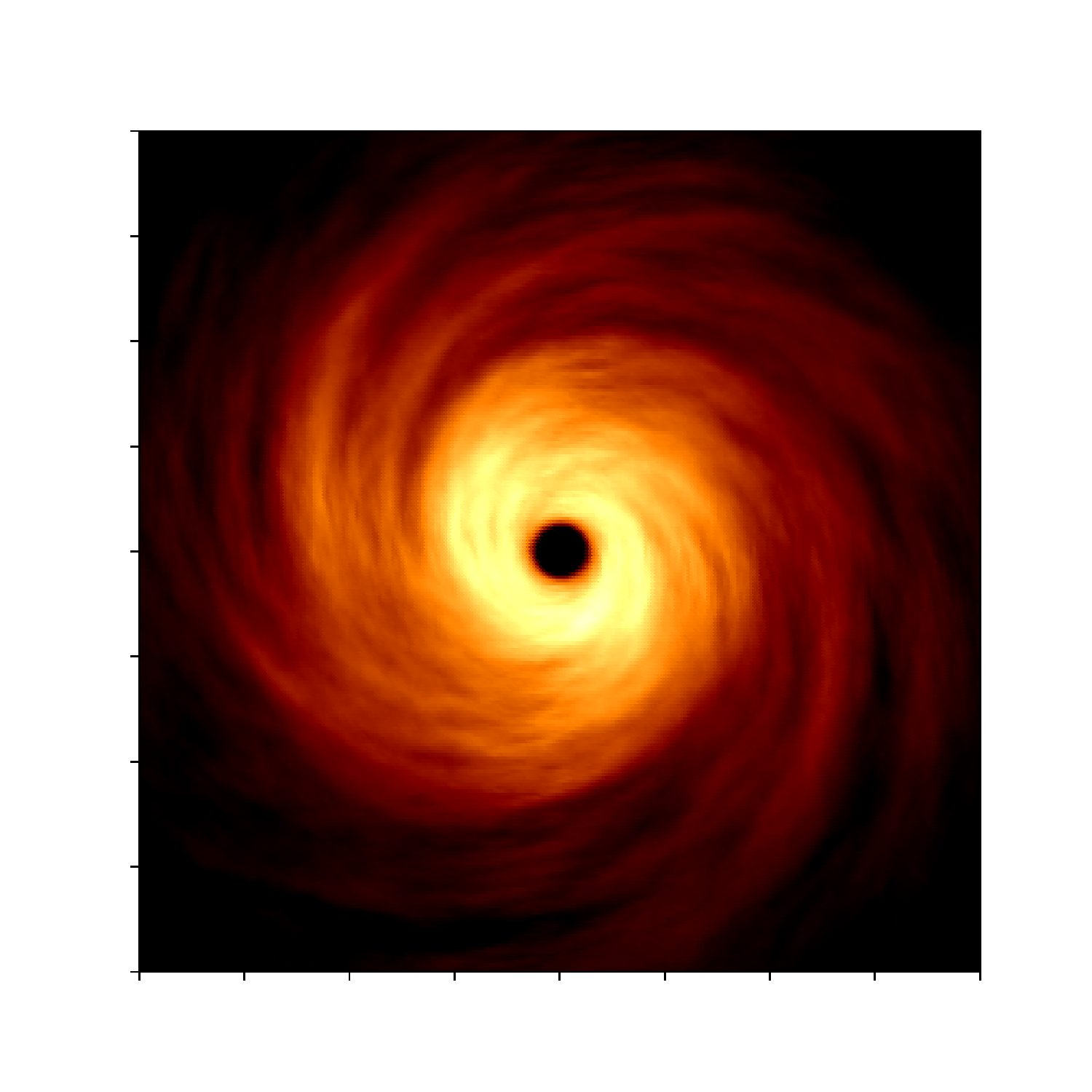}
		\includegraphics[width=0.32\linewidth]{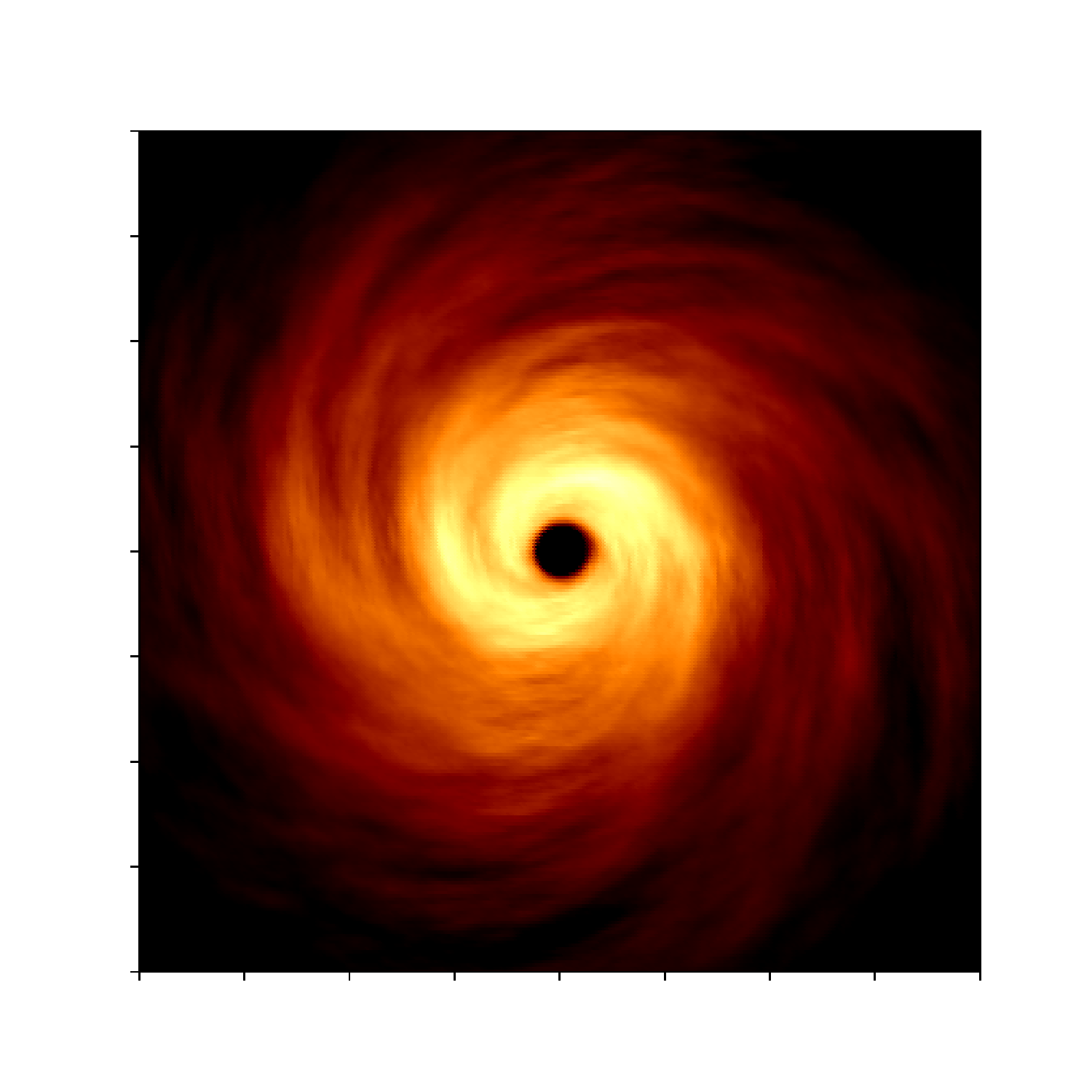}
		\includegraphics[width=0.32\linewidth]{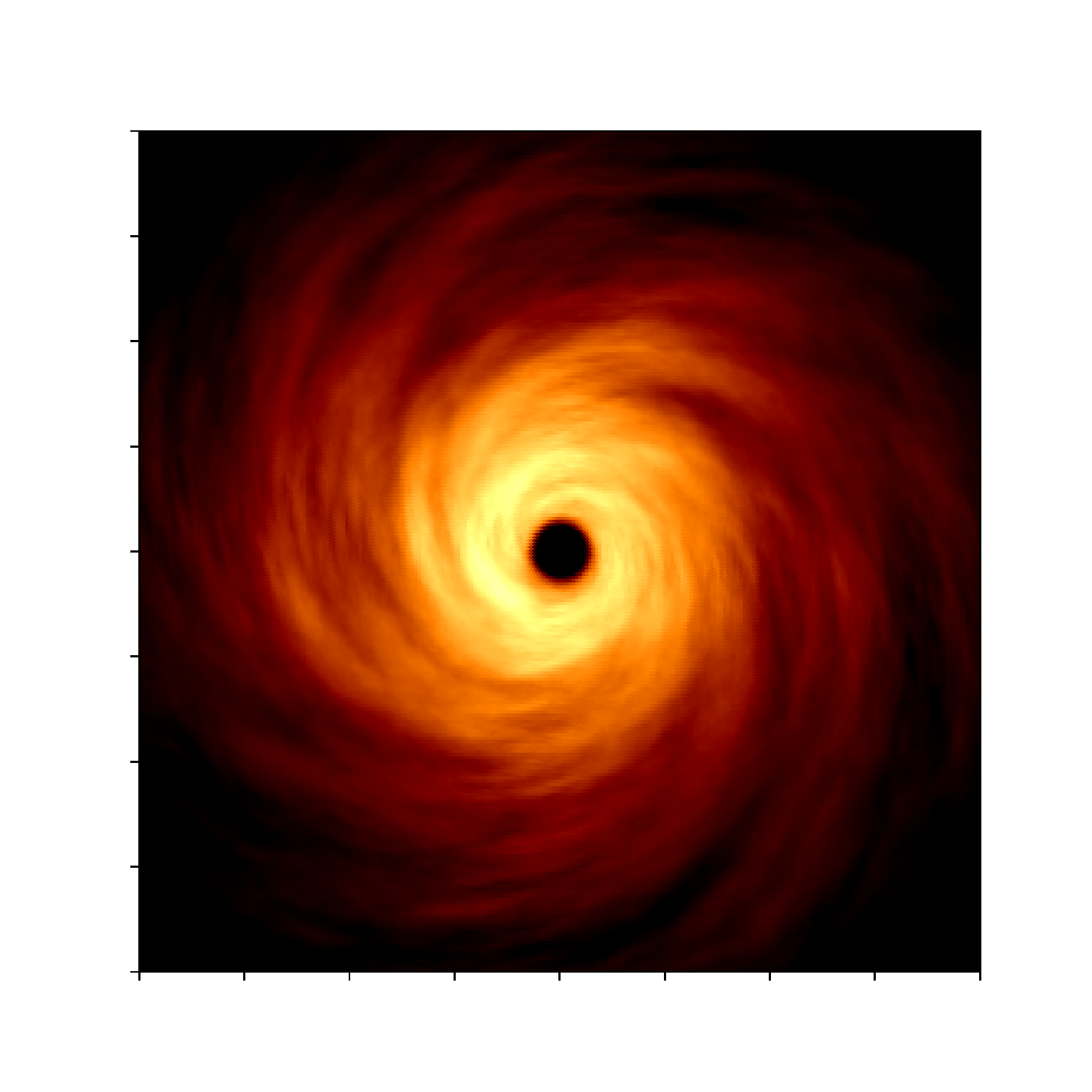}
\end{interactive}
\caption{A realization with differential rotation and position-dependent correlation lengths. The field, displayed in log scale, goes to zero in the center and near the edges due to the envelope function. As before, the appearance of advection is created by a spatiotemporal correlation that is applied along the velocity field, which for this realization is Keplerian. The position-dependent angle in the spatial correlation creates the spiral arm features. \label{fig:real}}
\end{figure}

Figure \ref{fig:real} shows the resulting non-stationary, anisotropic GRF with differential rotation and position-dependent correlation lengths.  The model was run on a Cartesian mesh in $x,y,t$ at a resolution of $(N_t, N_x, N_y) = (1024, 256, 256)$\added{, with periodic boundary conditions in time and Dirichlet boundary conditions in space. It was run} on an Intel Xeon 6140, and required less than two minutes to converge to a solution. The fluctuation field parameters are $\lambda_0(r) = \frac{2\pi}{\Omega_K}$, where $\Omega_K = \frac{1}{r^{3/2}}$, $\frac{\lambda_1}{r} = 5$, and $\frac{\lambda_2}{\lambda_1} = 0.1$.  The envelope has a single parameter $r_0$ that was set to 1/10th of the width of the grid ($\approx 26$ mesh points). 

The  code used to produce this example is publicly available at {\tt https://github.com/AFD-Illinois/inoisy}. 

\section{Application to Unresolved Disk Light Curves}
\label{sec:lightcurve}

AGN light curves are commonly modeled as a damped random walk  (Ornstein-Uhlenbeck process) \citep{kel09, mac10}. Ornstein-Uhlenbeck processes are the only nontrivial stochastic processes that are Gaussian, Markov, and stationary. The power spectrum of an Ornstein-Uhlenbeck process is
\begin{equation}
P(\omega) \propto (1 + (\tau \omega)^2)^{-1}
\end{equation}
where $\omega$ is the angular frequency and $\tau$ is a characteristic timescale. The power spectrum is flat (white noise) at low frequency and scales as $\omega^{-2}$ at high frequency. 
\added{Not all AGN light curves are well modeled by a damped random walk, however.  Some have steeper spectra at high frequency \citep[e.g.][]{smith18}.  Can we use the GRF model of \ref{sec:global} to produce a more general model that connects the local structure of the disk, encoded in $\mathbf{\Lambda}$, to the power spectrum of the light curve?}

A light curve is a time series constructed by integrating the source's fluctuating surface brightness over space at each instant.  For a GMRF realization $f(t,\bx)$\added{, where here $\bx$ is a 2D spatial vector, } the light curve is
\begin{equation}
L(t) = \int  g(\bx) \exp(\frac{f(t,\bx)}{n})\,d^2x
\end{equation}
where $g(\bx)$ is the envelope function.  

First, consider a simplified model calculation that shows that this integration can lead to interesting, nontrival results.  If $n$ is large then
\begin{equation}
L(t) \simeq L_0 + \int  g(\bx) f(t,\bx)\frac{1}{n} \,d^2x.
\end{equation}
where $L_0$ is the luminosity when $f = 0$.
To make the problem analytically tractable we use a stationary, isotropic GRF generated using the method of \S \ref{sec:examp} (Equation \ref{eq:spde} solved in 2 spatial dimensions plus time), with power spectrum 
\begin{equation}
P_f(\omega, \bk) = \frac{\sN}{(1+\omega^2/\omega_0^2+k^2/k_0^2)^2}.
\end{equation}
Here $\sN$ is a normalization constant.  

The power spectrum of the light curve $P_L(\omega)$ is the $\bm{k} = \bm{0}$ mode of the power spectrum $P_h(\omega,\bm{k})$ of $h(t,\bx)=f(t,\bx)g(\bx)/n$. Then, using a hat to denote the Fourier transform,
\begin{align}
P_L(\omega) &= P_h(\omega,\bm{0}) \nonumber \\
&= \reallywidehat{C(\Delta t,x,x') \cdot g(x) \cdot g(x')} \, \Big\rvert_{\bk=\bm{0}} \nonumber \\
&= \hat{C}*\hat{g}^2 \, \Big\rvert_{\bk=\bm{0}} \nonumber \\
&= \int P_f(\omega,\bm{k'}) \hat{g}^2 (\bk-\bk') \, d^2k' \, \Big\rvert_{\bk=\bm{0}} \nonumber \\
&= \int N \frac{\hat{g}^2 (-\bk')}{(1+\omega^2/\omega_0^2+k'^2/k_0^2)^2} \, d^2k'
\end{align}
Using a Gaussian envelope 
\begin{gather*}
g(\bx) = \frac{E}{2\pi\sigma^2}e^{-r^2/(2\sigma^2)} \\
\hat{g}(\bk)= \frac{E}{(2\pi)^2}e^{-k^2\sigma^2/2}	
\end{gather*}
the integrals can be done analytically and the light curve power spectrum is 
\begin{equation}
P_L(\omega) = \frac{E^2N}{2(2\pi)^3}\frac{k_0^2}{1+\tilde{\omega}^2}(1+Q^2e^{Q^2}\mathrm{Ei}(-Q^2))
\end{equation}
where $\tilde{\omega} = \omega/\omega_0$, $q^2 = (\sigma k_0)^2$, and $Q^2 = q^2 (1+\tilde\omega^2)$.

\begin{figure}[ht]
\gridline{\fig{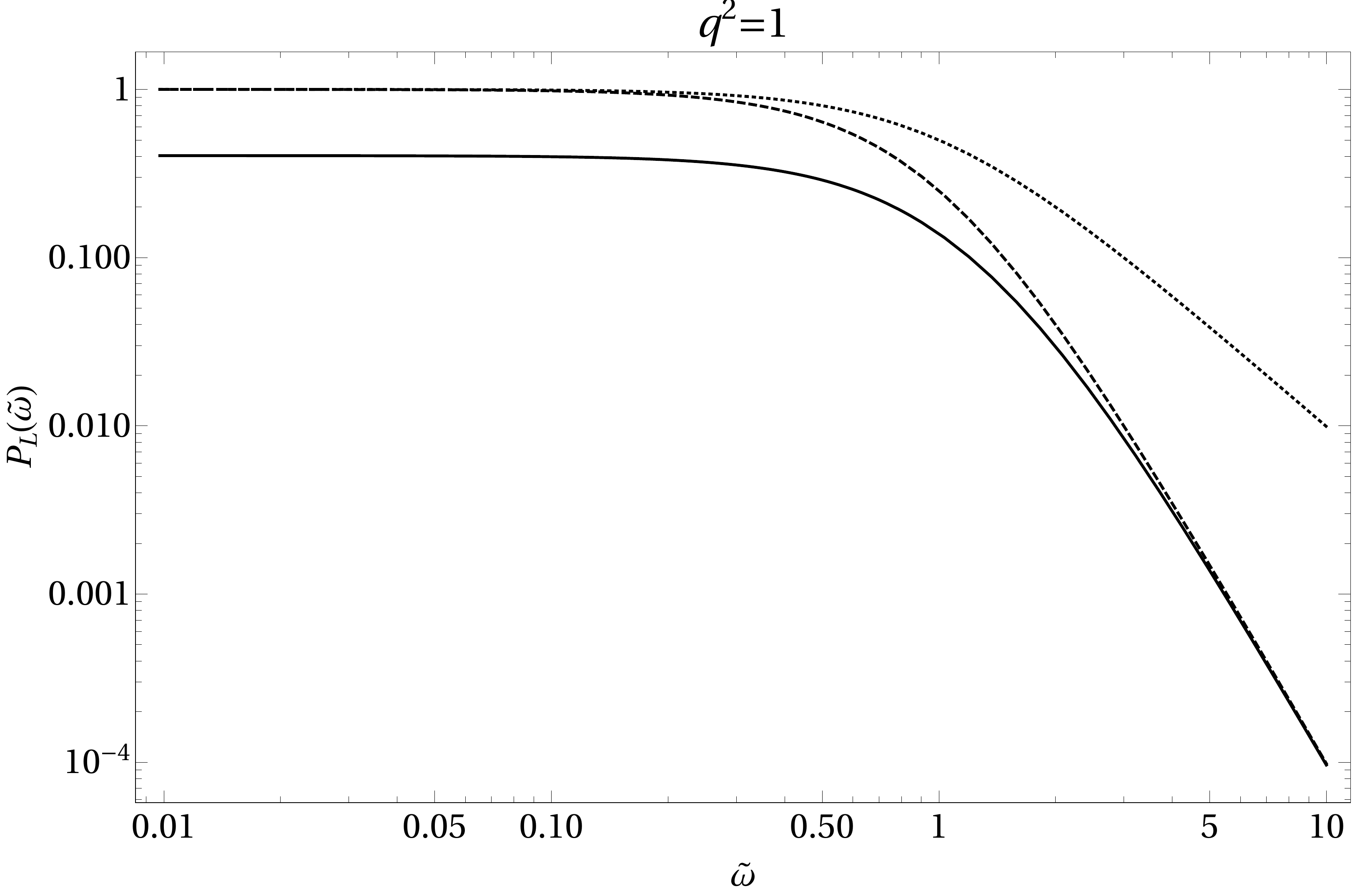}{0.49\textwidth}{(a)}
	\fig{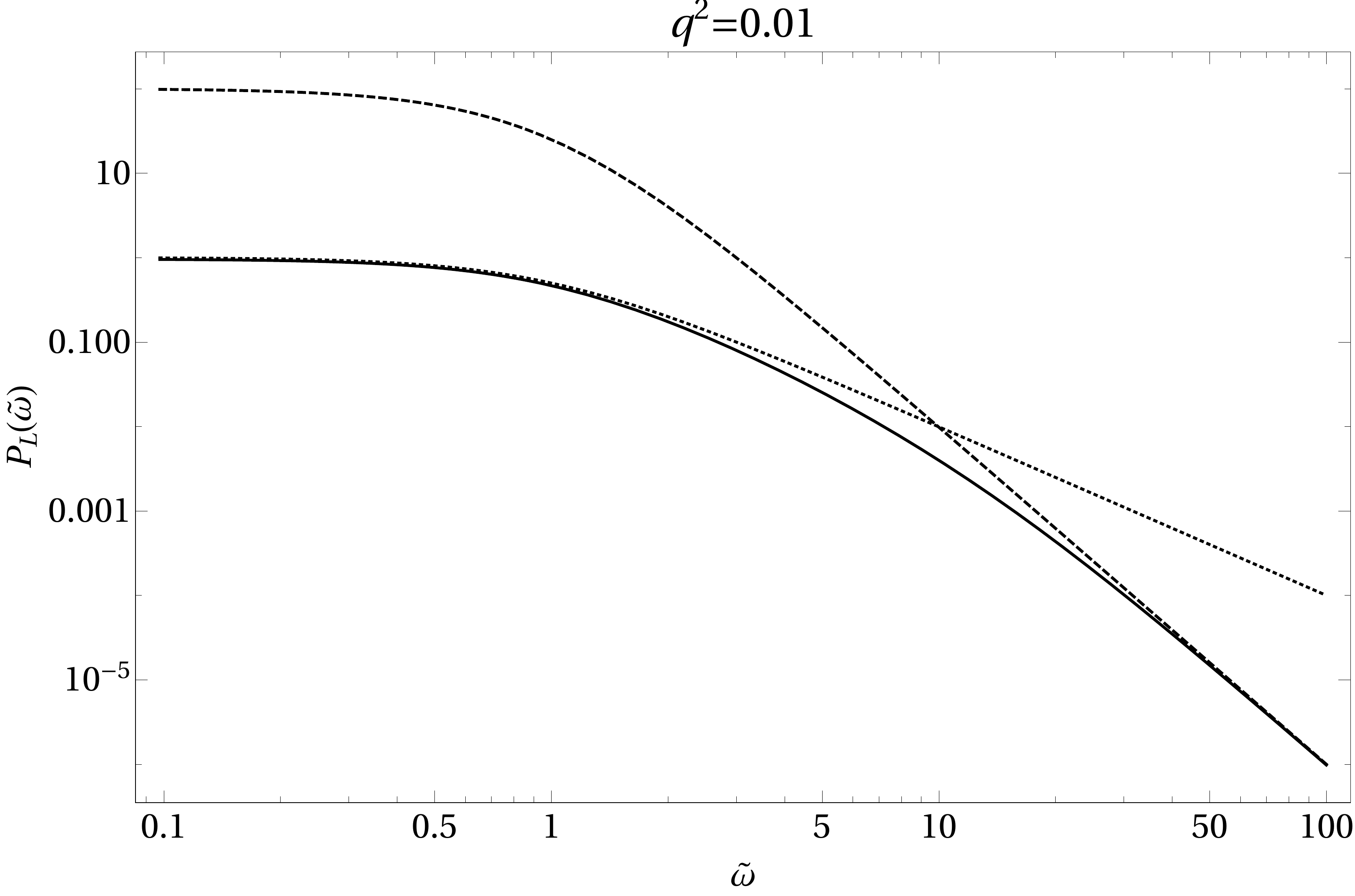}{0.49\textwidth}{(b)}}
\gridline{\fig{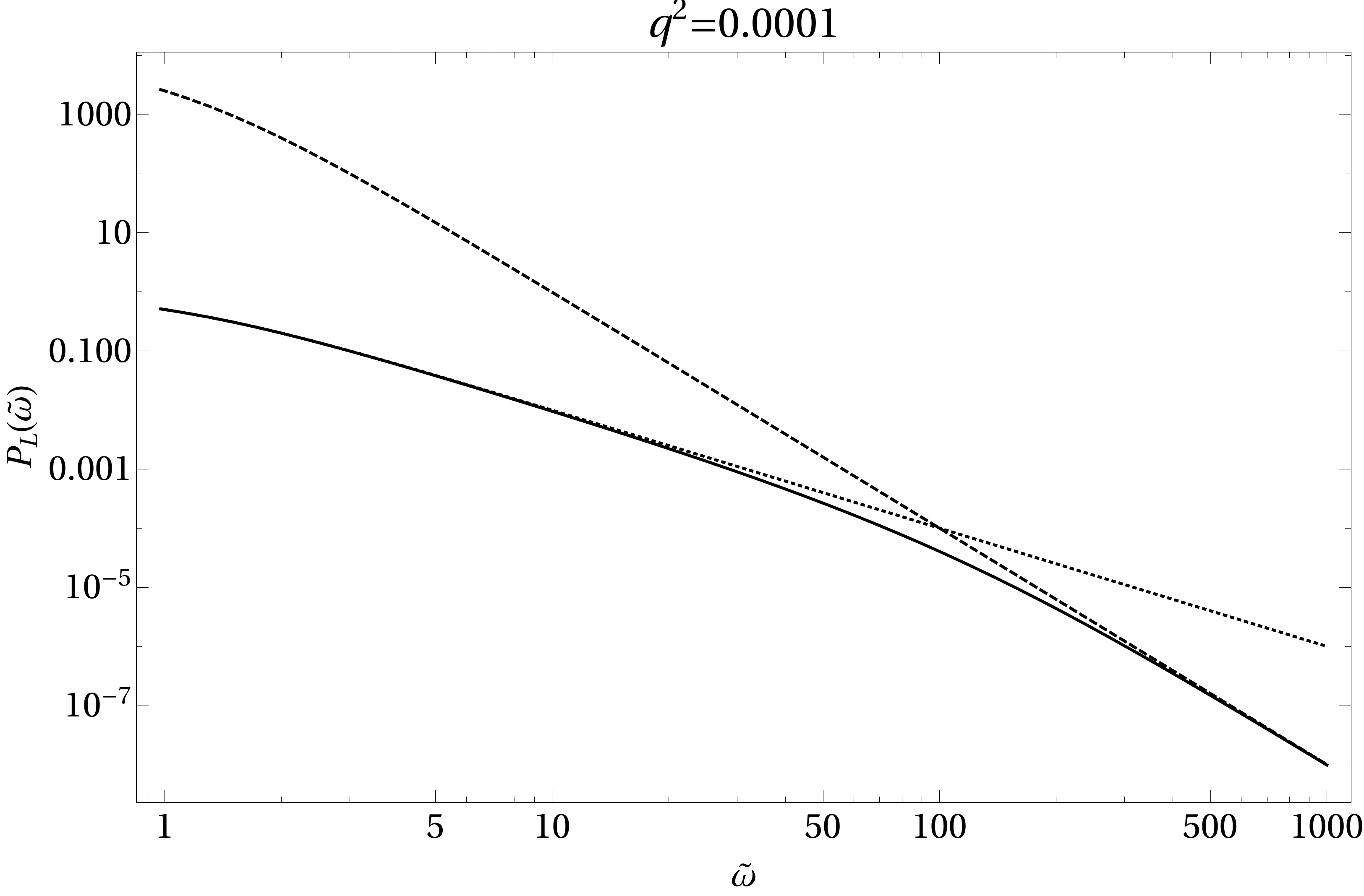}{0.49\textwidth}{(c)}}
\caption{Plot of $P_L(\tilde{\omega})$ (solid), $(1+\tilde{\omega}^2)^{-1}$ (dotted), and $\frac{1}{q^2}(1+\tilde{\omega}^2)^{-2}$ (dashed) for $q^2=1$ (a), 0.01 (b), and 0.0001 (c). }
\label{fig:power}
\end{figure}

When $\tilde\omega \ll 1$, $P_L$ approaches a constant, and when $q^2(1+\tilde\omega^2) \gg 1$, the power spectrum asymptotes to 
\begin{align}
P_L(\omega) &\rightarrow \frac{E^2N}{2(2\pi)^3}\frac{k_0^2}{1+\tilde{\omega}^2}Q^{-2} \nonumber \\
&\rightarrow \frac{E^2N}{2(2\pi)^3\sigma^2}\left(\frac{\omega}{\omega_0}\right)^{-4}
\end{align}
Thus, similar to the original power spectrum $P_f$, the light curve's power is flat at low frequencies and falls off as $\omega^{-4}$ at high frequencies. However, in the intermediate range where $\tilde\omega \gg 1$ but $Q \ll 1$, 
\begin{equation}
P_L(\omega) \rightarrow \frac{E^2N}{2(2\pi)^3}\left(\frac{\omega}{\omega_0}\right)^{-2}
\end{equation}
This intermediate regime occurs when $q^2=(\sigma k_0)^2=\left(\frac{2\pi\sigma}{\lambda}\right)^2$ is small, i.e. when the width of the envelope function is much smaller than the correlation length. Figure \ref{fig:power} shows the shape of $P_L$ with varying $q^2$. As $q^2$ becomes smaller, the intermediate regime becomes more prominent. 

The light curve power spectrum can also be calculated analytically for the more realistic envelope function 
\begin{equation}\label{eq:altenv}
    g(r) = \left(\frac{r}{\sigma}\right)^3e^{-r/\sigma}
\end{equation}
applied to the same homogeneous, isotropic GRF. Like the envelope used in \S\ref{sec:global}, this envelope is ring-like with a depression in the center, although it instead goes as $r^3$ at small $r$ and falls off exponentially at large $r$. 

The power spectrum of the light curve produced by \eqref{eq:altenv} has the form 
\begin{equation}
    P_L(\omega) \propto \frac{9 \pi q^4}{Q^3(1+Q)^{10}}\left[\sum_{i=0}^9a_iQ^i\right]
\end{equation}
where $a_i$ are integer coefficients and $\tilde{\omega}$, $q$, and $Q$ are defined as before. As for the Gaussian envelope, the power spectrum is constant for $\tilde{\omega} \ll 1$ and falls off as $\omega^{-4}$ at high frequencies, but in the intermediate regime where $\tilde{\omega} \gg 1$ but $Q \ll 1$ is small, the power spectrum falls as $\omega^{-3}$.  Once again, this regime is determined by $q$, the ratio of the envelope width to the correlation length. When the envelope is much larger than the correlation length, we recover the slope of $-4$, but as the envelope becomes narrower, the power spectrum becomes shallower. 

\begin{figure}[ht]
    \plottwo{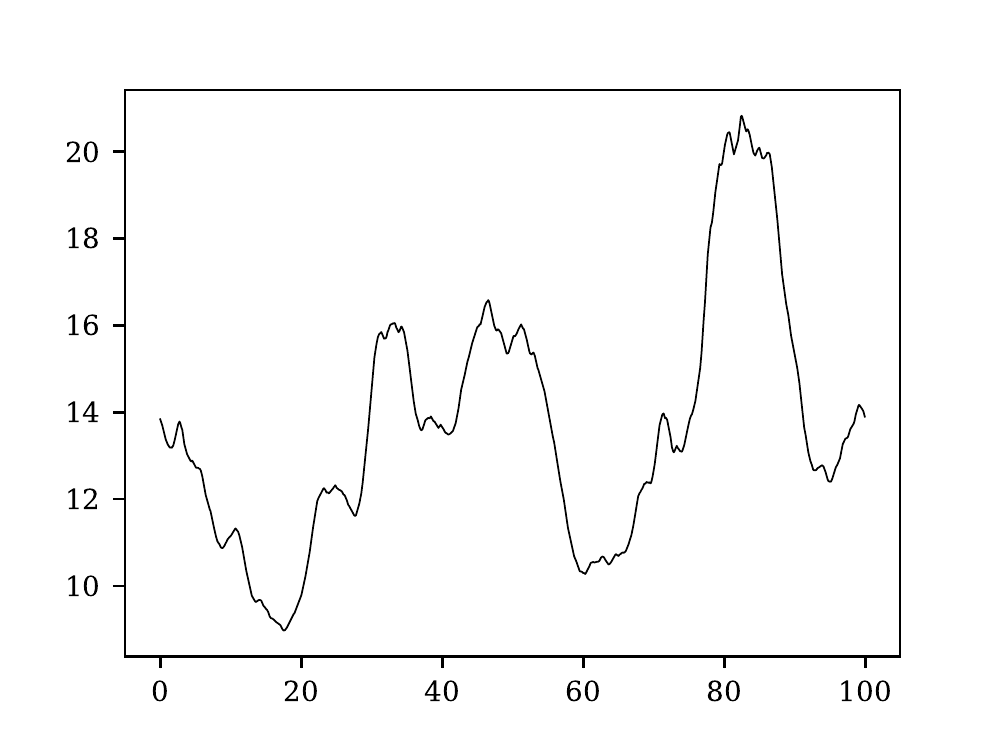}{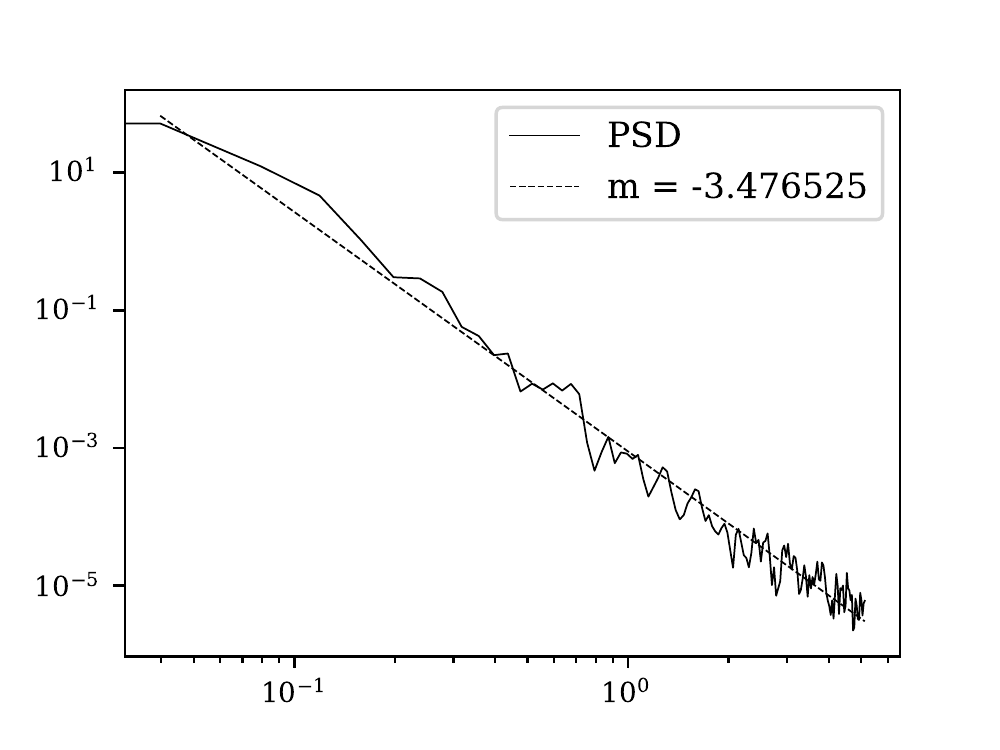}
    \caption{The light curve (left) and power spectrum (right) of the realization in Figure \ref{fig:real}. }
    \label{fig:realpow}
\end{figure}

Now consider the lightcurve of the global disk model generated in \S\ref{sec:global}.  Figure \ref{fig:realpow} shows the light curve and power spectrum of the realization shown in Figure \ref{fig:real}.  The slope of the power spectrum is shallower than $\omega^{-4}$ due to the combined effect of the envelope function and the inhomogeneity of the underlying GRF.  Because the fluctuation field is anisotropic and inhomogeneous, the preceding simple models do not apply, but the qualitative behavior is similar. 

\begin{figure}[ht]
    \plotone{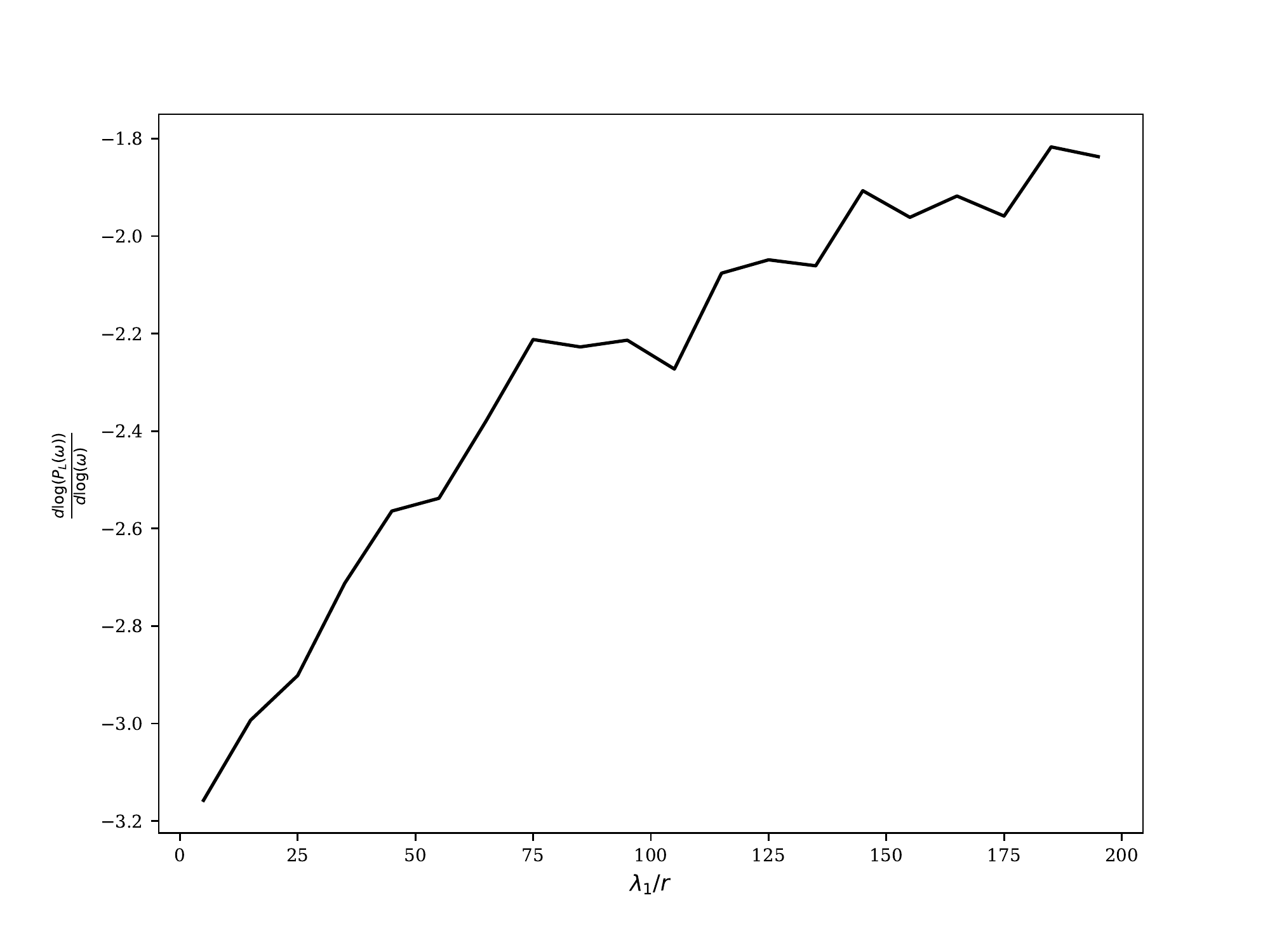}
    \caption{The slope of the power spectrum as a function of $\lambda_1/r$. The slope gets shallower as $\lambda_1/r$ increases. }
    \label{fig:pow_slope}
\end{figure}

It is interesting to ask whether the shape of the power spectrum, which is comparatively easy to observe, contains information about the underlying model.  Figure \ref{fig:pow_slope} shows how the high frequency slope of the lightcurve power spectrum for a family of realizations similar to \S\ref{sec:global}, with  $\lambda_1/r$ varying from 5 to 195.  The slope is averaged over 5 realizations. All other parameters are identical to \S\ref{sec:global}. Evidently the slope of the power spectrum exhibits behavior that is consistent with the simple models described above: the power spectrum becomes shallower as the ratio of the correlation length to the envelope width increases. 

\section{Summary}
\label{sec:summ}

We have explored a model in which surface brightness fluctuations on an astrophysical disk are treated as a Gaussian random field. Realizations of the anisotropic, inhomogeneous fluctuation field can be generated by solving the stochastic partial differential equation \ref{eq:spde}.  We provided pedagogical examples of anisotropic, inhomogeneous, and time-dependent anisotropic inhomogeneous fields in \S3.  The method requires that one specify an anisotropy tensor at every point in the domain.  The key, useful result is a simple parameterization of the anisotropy matrix \eqref{3dgrf}: one need only specify the anistropy and orientation of the covariance function in the two spatial dimensions, a velocity field, and a correlation time.  

We applied the method to realize time-dependent, resolved images of a statistical disk model.  The example shown in Figure \ref{fig:real} is difficult to distinguish from an animation of disk flow based on a physical simulation.  The method presented here enables an inexpensive statistical simulation of a disk (even in three spatial dimensions, as described briefly in the Appendix); the realization shown here was generated in a few minutes. 

[The physical inputs for the model are the correlation lengths and correlation times as a function of radius, which may be determined from simulations]

Our statistical disk models provide a complement to physical simulations.  Although physical simulations solve the governing equations and can therefore be predictive, they are subject to uncertainties related to physical and numerical approximations.  Statistical models -- if they can provide a good approximation to the simulations for particular parameter values -- enable one to treat the physical simulations as a point in a larger parameter space of models, and therefore provide a universe of models to test physical simulations against.  

We also used the statistical disk model to generate light curves for unresolved disks, and found that the models naturally produce an $f^{-2} - f^{-3}$ power spectrum.  This provides a means of connecting the space and time correlations of surface brightness fluctuations, which might be measured in a local model simulation, to the lightcurve.  

[ Evidently it is interesting to examine how to extract model parameters from a dataset (inference). ]

The technique used here seems likely to be useful elsewhere in astrophysics.  For example, in modeling turbulent fluctuations in the interstellar medium \citep{sal18}, providing realizations for turbulent inflow boundary conditions, or providing statistical models for any turbulent flow in which a mean flow field is known and for which a local covariance can be derived.   

\acknowledgments

This work was supported by the NSF grants  AST-1716327, OISE-1743747, and a Romano Professorial Scholarship.  \replaced{We thank Gil Holder and Aviad Levis for useful comments that improved the paper.}{We thank Gil Holder and Aviad Levis for their useful comments, and the referee for the helpful report that greatly improved the paper.}

\software{}

This paper made use of the {\tt hypre} library from {\tt https://github.com/hypre-space/hypre}.

\appendix

\section{Generalization to \texorpdfstring{$\nu \ne 1/2$}{nu != 1/2}}

In the models considered above we restricted attention to the Mat\'ern covariance with $\nu = 1/2$. A more general set of covariances in the Mat\'ern family can be found by solving the fractional SPDE 
\begin{equation}\label{eq:fracspde}
(1 - \lambda^2\nabla^2)^{\alpha/2}f(\bx) = \sN \lambda^{d/2} \sigma \, \sW (\bx)
\end{equation}
where $\sW$ is Gaussian white noise with unit variance and $\alpha = \nu + d/2$. The fractional differential operator $(\kappa^2 - \nabla^2)^{\alpha/2}$ is defined by its spectral properties: 
\begin{equation}\label{eq:matspec}
(1 - \lambda^2\nabla^2)^{\alpha/2} \phi_\bk = (1 + \lambda^2 \abs{\bk}^2)^{\alpha/2} \phi_\bk
\end{equation}
for any function $\phi$ for which the inverse Fourier transform of the right side is well defined.  Notice that for $\alpha = 2 m$, where $m$ is an integer, the differential operator in (\ref{eq:fracspde}) can be approximated by a finite difference and solved by conventional methods.  For example, the SPDE 
\begin{equation}\label{eq:delsq}
(1 + \lambda^2\nabla^2)^2 f(\bx) = \sN \lambda^{d/2} \sigma \sW (\bx)
\end{equation}
has
\begin{equation}\label{eq:delsqpsp}
P_k = \frac{\sN^2 \lambda^d \sigma^2}{(1 + \lambda^2 k^2)^4}.
\end{equation}
One motivation for considering a higher order model like this is for modeling a time-dependent process in three spatial dimensions ($d = 4$);  for $\nu = 1/2$ and $d > 3$ the variance of the GRF diverges, and the $\nu = 1/2$ model is unsatisfactory \added{, although this defect can be repaired by replacing the white noise process $\sW$ by a red noise process, possibly realized as the solution to a separate SPDE.}

Notice that any finite difference operator with a compact  stencil will have the Markov property, and the associated SPDE generates a GRF.   

\section{Alternative Approach to Time Dependence}

In \S \ref{sec:examp} we introduced time dependence by treating the time coordinate on the same footing as the space coordinates, and orienting one axis of the correlation ellipse along a velocity vector.  Here we consider a distinct procedure based on the SPDE
\begin{equation}
    \left(1 + \tau \left(\frac{\del}{\del t} + \bm{v} \cdot \nabla \right) - \lambda^2 \nabla^2
    \right) f(\bx) =
    A \sW(\bx)
\end{equation}
where $\tau$ is a characteristic timescale and $A$ is a constant.  This SPDE was proposed by \cite{lin11} using slightly different notation.  The power spectrum is
\begin{equation}
    P_{\omega,\bk} \propto 
    \left(\tau (\omega - \bk \cdot \bv)^2 + (1 + \lambda^2 k^2)^2 \right)^{-1}.
\end{equation}
from which one can see (because the integral over $\omega, \bk$ diverges in two spatial dimensions) that the pointwise variance of this model is formally infinite.  The variance of the associated GMRF on a finite grid is not, however.  We have implemented this form in the {\tt noisy} code, available at {\tt https://github.com/AFD-Illinois/inoisyB}.  

Another SPDE that produces a field that is equivalent to the example in \S \ref{sec:examp} is
\begin{equation}
    \left(1 - \tau^2 \left(\frac{\del}{\del t} + \bv \cdot \nabla \right)^2 - \lambda^2 \nabla^2
    \right) f(\bx) = \sN \sigma \lambda^{d/2} \tau^{1/2} \sW(\bx).
\end{equation}
where $d$ is the number of spatial dimensions.  
The power spectrum is
\begin{equation}
    P_{\omega,\bk} \propto 
    \left(\tau^2 (\omega - \bk \cdot \bv)^2 + 1 + \lambda^2 k^2)^2 \right)^{-1}.
\end{equation}
Evidently the associated GRF has finite variance in two spatial dimensions, but not in three.

\end{document}